\documentclass[aps,prl,amsfonts,amssymb,epsfig,graphics,showpacs,10pt,twocolumn]{revtex4}

\usepackage{graphicx}
\usepackage{epsfig}

\begin{document}

\def\ve{\varepsilon}
\def\cG{{\cal G}}
\def\us{\underline{\sigma}}
\def\expect{{\mathbb E}}
\def\<{\langle}
\def\>{\rangle}

\date{\today}

\title{From Large Scale Rearrangements to Mode Coupling Phenomenology}
\author{ A.~Montanari$^{\,1}$, G.~Semerjian$^{\,2}$}
\affiliation{$^{1\,}$Laboratoire de Physique Th\'{e}orique de l'Ecole Normale
  Sup\'{e}rieure, CNRS-UMR 8549}
\affiliation{$^{2\,}$
Dipartimento di Fisica, INFM (UdR and SMC centre), 
Universit\`{a} di Roma ``La Sapienza''}
\pacs{75.50.Lk (Spin glasses), 64.70.Pf (Glass transitions), 
89.20.Ff (Computer science)}

\begin{abstract} 
We consider the equilibrium dynamics of Ising spin models with multi-spin 
interactions on sparse random graphs (Bethe lattices). 
Such models undergo a mean field glass transition
upon increasing the graph connectivity or lowering the temperature.
Focusing on the low temperature limit, we identify the large scale 
rearrangements responsible for the dynamical slowing-down near the transition.
We are able to characterize exactly the critical dynamics by 
analyzing the statistical properties of such rearrangements. 
We obtain a precise crossover description of the role of
activation at the transition.
Our approach can be generalized to a large variety of glassy models
on sparse random graphs, ranging from satisfiability to kinetically
constrained models.
\end{abstract}

\maketitle

Understanding the slowing down of relaxational dynamics in 
glass-forming liquids is an important open problem in statistical physics.
Two points of view have been developed in the last years. 
Mode coupling theory (MCT)~\cite{MCT} is based upon a ``self-consistent''
closure of dynamical equations, and predicts an ergodicity-breaking 
transition at temperature  $T_{\rm d}$.
It was later shown that MCT is exact for a class of fully-connected
(FC) mean-field models,
the dynamical phase transition (DPT) being related to the proliferation 
of metastable states~\cite{KiTh,MCA}.

According to an alternative point of view, the 
dynamical slowing down in supercooled liquids can be traced 
back to the increasing cooperativity of the dynamics~\cite{BeGaJPB}. 

It is a recent discovery that MCT implies diverging correlations as 
the  $T_{\rm d}$ is approached~\cite{BB}. 
This hints at a possible convergence among the above points of view, 
and may lead to universal predictions. 
However, the relation
between a diverging correlation length and dynamical slowing-down remains
qualitative. This paper aims at filling this gap.
By analyzing a particular case, we will show that a detailed picture 
of the critical dynamics can be obtained through the analysis of
highly correlated regions whose size diverges at the transition. 
Several features of MCT are recovered, despite no exact
set of MCT equations holds in the system considered here~\cite{SeCuMo}.

A further source of motivation comes from the discovery
that several {\em ensembles} of hard optimization problems, such as 
satisfiability and coloring, 
undergo a mean-field DPT~\cite{notrerevue}. 
An interesting question in this 
context is: how much time a Monte Carlo (MC) algorithm
needs for sampling a low-cost configuration of the problem. 
Furthermore, a phase transition of the same type,
is found in a large variety of other models
on random graphs, from kinetically constrained models, to
rigidity percolation~\cite{RigidityAndCo}.

Finally, there has been a lot of interest in the
role of `activated processes' in glasses~\cite{Kob}. 
In (spherical) FC models free energy barriers vanish above 
$T_{\rm d}$ while they are extensive (in the system size $N$) below.
Activation does not play any role: it is not necessary above $T_{\rm d}$,
and it is ineffective below. Schematic MCT does not include activation and
is exact for such models, predicting a sharp transition. 
This picture can be modified in two ways:
(A) Requiring barriers to stay finite below $T_{\rm d}$. In finite-dimensional
models, this is a consequence of nucleation effects. Activation
over such barriers induces a finite time scale for ergodicity restoration, 
and a smearing of the transition. (B) Introducing finite barriers above 
$T_{\rm d}$ while keeping extensive ones below. 
This is the scenario in \emph{diluted} mean field models,
such as the  one treated below. While the transition remains sharp,
the presence of new (activated) relaxation mechanisms
may change the slowing down as $T\downarrow T_{\rm d}$,
(e.g. the critical exponents).

In this paper we address the problem (B) above. Since in any realistic 
(finite-dimensional) model crossing over finite-energy barriers exists
at any temperature, it is important to understand if this may modify
some key  predictions of MCT.
In particular, we shall consider 
a regime in which  the distinction between activated and non-activated 
processes has a precise mathematical sense,
i.e. in the neighborhood of a $T=0$ bicritical point.

Our approach is to study  a class of large scale rearrangements (LSR)
which we expect to control the slow dynamics.
Let us begin by providing a loose description of the main ideas on the 
example of a particle system. Consider a 
low-$T$~\footnote{By this we mean close or below the MCT
transition.} equilibrium configuration and 
focus on a particular molecule at position $x_i$.
Now impose a displacement $\delta x_i$ (a few intermolecular distances) on 
this molecule and ask what is the {\em minimum} number $n_i$ of 
other molecules  which must be moved for this displacement to be possible.
The {\em minimum} energy barrier $b_i$ to be overcome is a second 
important property of the displacement $\delta x_i$.
A glassy state is characterized by very large sizes $n_i$'s 
and barriers $b_i$'s (diverging at a sharp DPT)
leading in turn to large relaxation times.
A third quantity is defined by allowing
all the molecules within a distance $\ell$ around $x_i$ to be 
moved. Let $\ell_i$ be the minimum $\ell$ such that the 
displacement $\delta x_i$ can be performed. 
It turns out that $n_i\ll \ell_i^{d}$ (at high enough dimension $d$). 
Remarkably, molecular dynamics simulations~\cite{rearr} found
string-like motions in glass-forming liquids.

Such notions can be completely precised in specific models. 
In this paper we focus on Ising models with $p$-spin ($p\ge 3$) interactions:
\begin{equation}
H(\sigma)=\frac{1}{2} 
\sum_{(i_1\dots i_p)\in\cG} (1-J_{i_1 \dots i_p} \sigma_{i_1} 
\cdots \sigma_{i_p}) \, ,\label{Hamiltonian}
\end{equation}
Here $\sigma_i=\pm 1$ are $N$ Ising spins, $\cG$ is a set of
$M=N\gamma$ $p$-uples of indices, and $J_{i_1 \dots i_p}$ 
are quenched couplings taking values $\pm 1$ with equal probability.
The above Hamiltonian counts the number of violated
constraints $ \sigma_{i_1} \cdots \sigma_{i_p} =J_{i_1\dots i_p}$.
This model is known in
computer science as XORSAT~\cite{XOR_CS}.

\begin{figure}
\includegraphics[width=7.5cm]{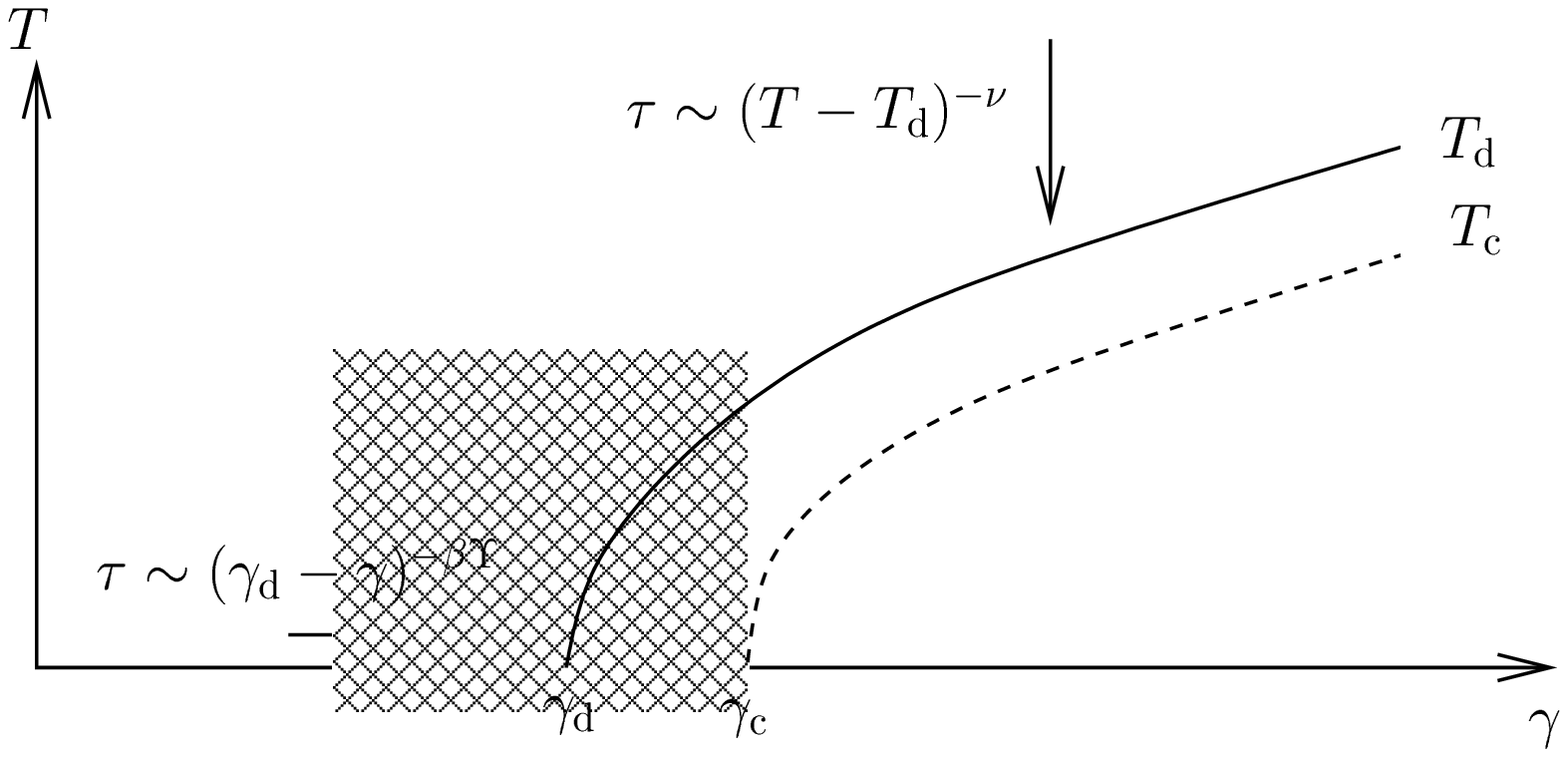}
\caption{Schematic view of the phase diagram of the diluted $p$-spin model,
with its dynamical and static transition line. See the text for details on
the two type of divergences encountered at the dynamical transition. 
The scaling hypothesis of Eq.~(\ref{eq:TemperatureScaling}) is expected to 
hold in the shaded region.}\vspace{-0.4cm}
\label{phasediag}
\end{figure}

A mean field version of this model is obtained by taking the $M$
interacting $p$-uples of sites $\{(i_1\dots i_p)\}$ to be quenched 
random variables uniformly distributed in  the set of 
$N \choose p$ possible $p$-uples~\cite{XOR}.
In the limit $N\to\infty$, the number of interactions a given 
spin belongs to (its connectivity) is a Poisson random variable with parameter 
$p\gamma$. Moreover, the shortest loop through such a spin is (typically)
of order $\log_{p\gamma} N$. 
The phase diagram is sketched in Fig.~\ref{phasediag}.
Two regimes have attracted a particular attention. In the ``fully-connected''
limit $\gamma\to\infty$, $T\propto \sqrt{\gamma}$, both statics and
dynamics can be treated analytically, showing a typical MCT transition. 
The relaxation time diverges as $|T-T_{\rm d}|^{-\nu}$, 
while a true thermodynamic transition takes place at $T_{\rm c}$.

In the $T\downarrow 0$, finite $\gamma$, limit, probabilistic  
methods can be used to show that  zero-energy ground states 
with finite entropy density exist for $\gamma<\gamma_{\rm c}$.
The set of ground states gets splitted in an exponential number of clusters
with extensive Hamming distance (number of spins with different value)
separating them for 
$\gamma_{\rm d}<\gamma<\gamma_{\rm c}$~\cite{XOR_12}. 
A finite fraction of the spins 
does not vary within a particular cluster, while
they change when passing from a cluster to the other. 
At $\gamma_{\rm d}$ the fraction of ``frozen'' spins jumps from $0$
to a finite value $\phi$. For instance $\gamma_{\rm d}\approx 0.81847$,
$\gamma_{\rm c}\approx 0.91793$, and $\phi\approx 0.71533$ when $p=3$. 
No exact result exists for the dynamics in this regime~\cite{SeCuMo}.

In the region $T<T_{\rm d}(\gamma)$, 
the Gibbs measure splits
into an exponential number of pure states separated by barriers 
of order $N$. However this implies little (if anything) on the 
correlation time behavior as $T\downarrow T_{\rm d}(\gamma)$. 
In agreement with our general philosophy, we shall analyze the  
dynamics in terms of LSR's
as the transition line $T_{\rm d}(\gamma)$ is approached.
The discussion below concerns any single-spin-flip Markov dynamics
satisfying detailed balance. 

Notice that a diverging length cannot be defined through 
a standard spin-glass correlation function, which remains short-ranged
at $T_{\rm d}(\gamma)$. In order to overcome this problem,
consider any temperature 
$T>T_{\rm d}(\gamma)$ and fix a reference thermalized configuration
$\us^{(0)}$, a site $i$, and a length $\ell$. Denote by 
$\<\sigma_i\>_{\ell}$ the Boltzmann average of $\sigma_i$ under the 
boundary condition  $\sigma_j=\sigma^{(0)}_j$ for any site $j$ at a distance 
larger than $\ell$ from $i$. Define $\ell_i(\ve)$ to be the smallest value of
$\ell$ such that  $\sigma^{(0)}_i\<\sigma_i\>\le \ve$ ($\ve$ being a small 
fixed number). 
A standard recursive calculation
yields $\ell_i(\ve)\sim (T-T_{\rm d}(\gamma))^{-1/2}$ as
$T\downarrow T_{\rm d}(\gamma)$. A coupling 
argument from probability theory can be used~\cite{OurFuture,Proba}
to show that this implies a correlation time 
$\tau\gtrsim (T-T_{\rm d}(\gamma))^{-1/2}$.

The above argument displays clearly the relation between 
length and time scales. However the estimate for $\tau$
is not tight (the exponent is incorrect). As we shall see next, a highly 
refined picture
can be obtained as $T\to 0$ in the ``liquid'' phase $\gamma<\gamma_{\rm d}$.
In this regime, the system will spend most of its time in quasi-ground states.
For the sake of the argument, assume that it is in fact in a ground state
and focus on a particular spin $\sigma_i$. The leading mechanism
for $\sigma_i$ to relax consists in a trajectory in phase space which brings 
the system to a new ground state with a reversed value of $\sigma_i$.
Let $F_i$ be the set of reversed spins between
these two ground states. This set must contain $i$ and, for each interaction 
$(i_1\dots i_p)\in \cG$, an even number of these $p$ sites.
As will be clear from the following, we can restrict in fact to those
$F_i$'s which are connected, and such that, for each interaction 
$(i_1\dots i_p)\in \cG$,
either two or none of the sites belong to $F_i$.
In the present context, we call $F_i$ a {\em rearrangement} for 
the spin $\sigma_i$.

\begin{figure}
\includegraphics[width=6.5cm]{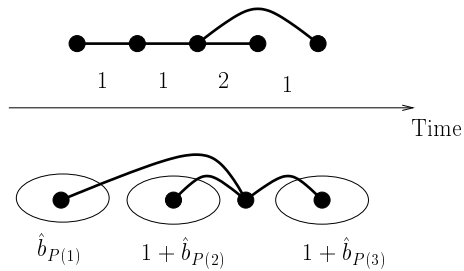}
\caption{Top: an optimal ordering for this rearrangement $F$. The numbers
indicate the energy along the trajectory, here $b(F)=2$. Bottom: example of
the recursive construction for $l=3$, $j=2$. Bubbles stand for generic
rooted sub-trees. The numbers give the maximal energy in each epoch.}
\vspace{-0.4cm}
\label{fig_disjoint}
\end{figure}
Each $F_i$ can be assigned a barrier $b(F_i)$, defined as
the minimum over all (single-spin-flip) paths in configuration
space which flip the spins of $F_i$, of the maximum energy 
along the path. Assume 
that each spin in $F_i$ is flipped only once:
paths are thus defined by an ordering of the flipped variables.
A relaxation time for $\sigma_i$  can be defined 
by considering the correlation function 
$C_i(t) = \<\sigma_i(t)\sigma_i(0)\>$ and requiring $C_i(\tau_i) = \ve$
for some fixed $\ve$ (in the following $\ve = 1/2$). 
At low temperature, Arrhenius law implies $\tau_i \sim \exp[\beta b_i]$,
with $b_i = \min_{F_i}b(F_i)$.

Computing $b_i$ requires optimizing both over the choice of the 
rearrangement $F_i$ (the set of spins to be flipped), and over
the paths in configuration space (the flipping order). Consider 
this second task for a given rearrangement $F_i$. 
Form the graph with vertices representing the spins of $F_i$, and
links between spins belonging to a common interaction in $\cG$.
If $F_i$ stays finite in the thermodynamic limit, 
this graph is a tree rooted at $i$. 
If one draws this tree with vertices placed 
on an horizontal axis according to the order in which the 
spins are flipped, the energy of the system at a certain 
point of the trajectory is simply the number of links drawn 
above this point, cf. Fig.~\ref{fig_disjoint}.
This ordering problem  is studied in graph 
theory as \emph{minimal cutwidth}~\cite{Yanna,Lengauer}. 
 
A simple (and essentially optimal~\cite{OurFuture})
strategy to construct recursively such an ordering
is the following. 
Assume that the site $i=0$ has $l$ neighbors $\{1\dots l\}\equiv[1,l]$, 
each one being 
the root of a sub-tree. Then choose a sequential ordering
of the sub-trees, i.e. a permutation $P$ of 
$[1,l]$, and an integer $j \in [0,l]$. Flip all the variables of the 
sub-tree $P_1$,
then do the same on the tree $P_2$, and so on 
until $P_j$, then flip the  $\sigma_0$, and finally flip the
spins in the sub-trees $P_{j+1},\dots,P_{l}$. As ``proved'' in 
Fig. \ref{fig_disjoint}, this construction implies a recursion on the 
energy barriers
\begin{eqnarray}
b_0 & = & \min_{P,j} \max [\;\hat{b}_{P_{1}}, 1 +\hat{b}_{P_{2}}, \dots,
j-1 +\hat{b}_{P_{j}}, \label{eq_disjoint_0}
\\ &&\phantom{\min_{P,j} \max[\;} l-j-1 +r+\hat{b}_{P_{j+1}}, \dots,
r+\hat{b}_{P_{l}}\;]\, ,\nonumber
\end{eqnarray}
\vspace{-0.4cm}\\
with $r=0$.
The $\hat{b}_i$'s are ``modified barriers''  which obey a
similar recursion: $\hat{b}_0$
is given by Eq. (\ref{eq_disjoint_0}) with $r=1$.

The computation of the barrier $b_i$ for the
$p$-spin interaction problem still involves an optimal choice of
the rearrangement $F_i$. 
This step can also be performed recursively: starting from
the root $i$, one chooses in each interaction around it the variable $a$
(among the $p-1$ distinct from $i$) which minimizes $\hat{b}_a$. Repeating
this step one obtains an admissible rearrangement $F_i$ for 
$\sigma_i$, with a minimal value of $b(F_i)$.
Remarkably, this scheme can be efficiently implemented on a given 
sample~\cite{OurFuture}.

\begin{figure}
\includegraphics[width=7.5cm]{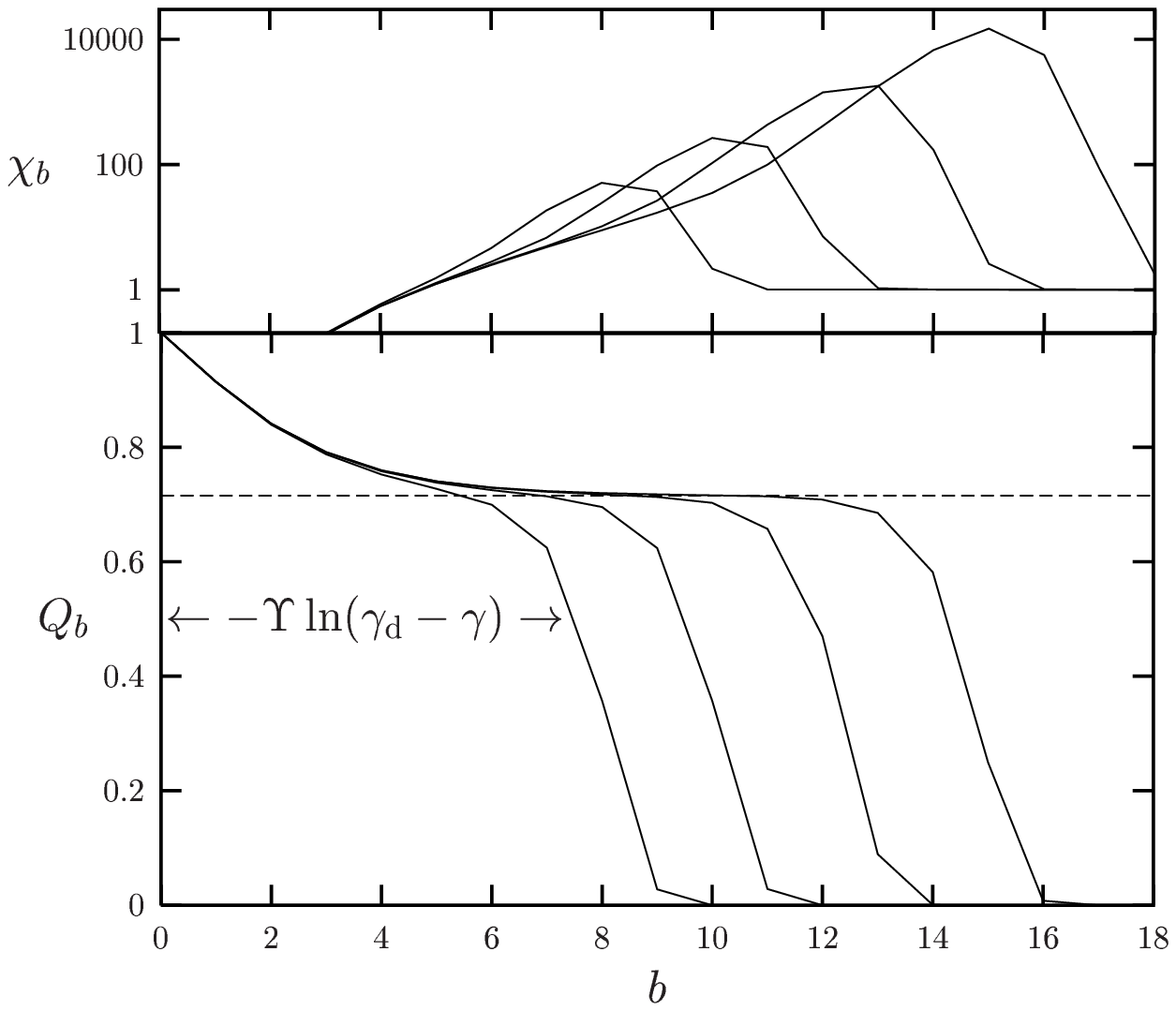}\vspace{-0.4cm}

\caption{Bottom: integrated law for the distribution of barrier heights.
Top: geometrical susceptibility.}\vspace{-0.3cm}

\label{fig_pih}
\end{figure}

If we consider the ensemble of random hypergraphs ${\cal G}$
described above, the  barriers $b_i$ become themselves random variables, 
and Eq.~(\ref{eq_disjoint_0}) acquires a distributional meaning.
The law $Q_b \equiv  {\rm Prob} [b_i\ge b]$ can be computed
numerically and is plotted in Fig.~\ref{fig_pih} for a few values of 
$\gamma$ approaching $\gamma_{\rm d}$. Notice that $Q_b$
has an immediate physical interpretation in terms of the global correlation
function $C(t) = \frac{1}{N}\sum_i \<\sigma_i(0)\sigma_i(t)\>$. 
At time $t$ only those sites with $b_i\gtrsim T\log t$ contribute to the
correlation function. We thus have 
$\lim_{\beta\to\infty}C(e^{\beta (b-\delta)})= Q_b$ for any $0<\delta<1$.
The critical behavior of $Q_b$ can be solved analytically.
As $\gamma$ approaches $\gamma_{\rm d}$ a plateau develops at height $\phi$:
a fraction $\phi$ of the spins have ``large'' barriers (are freezing), 
while the other ones have ``small'' barriers. 
As the plateau is approached (left) one has 
$Q_b\simeq \phi + C_- e^{-\omega_ab}$
($Q_b\simeq \phi - C_+ e^{\omega_bb}$) with
$\omega_{a/b}$ the positive solutions of the equations
$2e^{\omega_a} - e^{2\omega_a} = 2e^{-\omega_b} - e^{-2\omega_b} = \lambda$, 
and $\lambda$  a $p$-dependent parameter. 
Finally the scale of the large  barriers diverges as 
$b_{\rm slow}\simeq - \Upsilon\log(\gamma_{\rm d}-\gamma)$, with
$\Upsilon = 1/(2\omega_a) + 1/(2\omega_b)$. For instance, if $p=3$ 
we get $\omega_a\approx 0.57432$, $\omega_b \approx 1.49574$,  and 
$\Upsilon \approx 1.20488$. The reader will notice the 
close parallel with the behavior of correlation functions in MCT, 
with some definite differences: here the divergence is logarithmic rather than
power-law; the exponents are no longer fixed by a transcendental relation
(see below),
but rather through the above quadratic equations for $e^{\omega_{a/b}}$.

\begin{figure}
\includegraphics[width=8.cm]{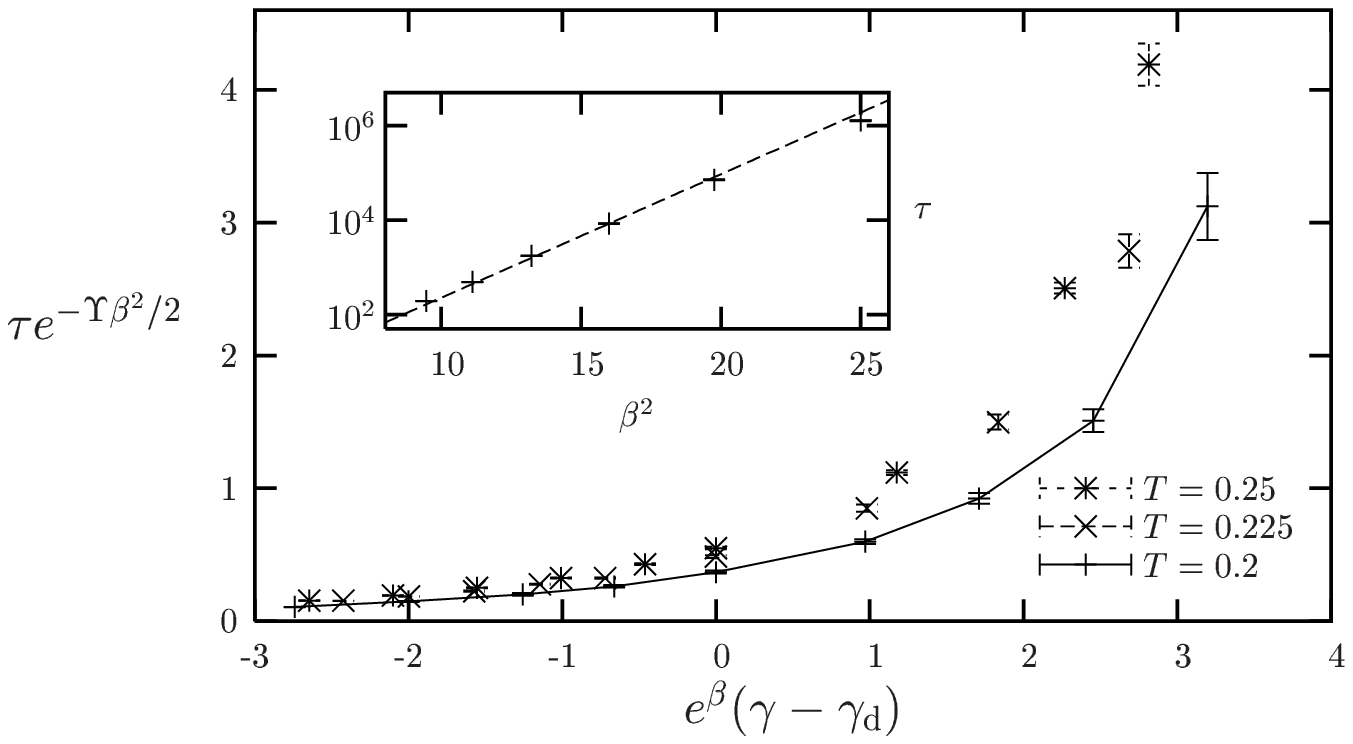}\vspace{-0.4cm}

\caption{The scaling function of Eq.~(\ref{eq:TemperatureScaling})
(notice the large corrections to scaling for 
$\gamma>\gamma_{\rm d}$). 
Inset: super-Arrhenius behavior at $\gamma_d$.
}\vspace{-0.3cm}
\label{fig_scaling}
\end{figure}

Arrhenius law  yields a correlation time diverging as  
$\tau\simeq (\gamma_{\rm d} - \gamma)^{-\beta \Upsilon}$  if the 
$\gamma\to\gamma_{\rm d}$ limit is taken {\em after} $T\to 0$.
Inspired by the crossover behavior in diluted ferromagnets~\cite{Henley}, 
we conjecture the following scaling form to 
hold if  $\gamma\to\gamma_{\rm d}$ together with $T\to 0$:
\begin{eqnarray}
\tau(\beta,\gamma)
\simeq e^{\frac{\Upsilon\beta^2}{2}}f(e^{\beta}(\gamma-\gamma_{\rm d}))\, .
\label{eq:TemperatureScaling}
\end{eqnarray}
This summarizes the above findings, as well as the low-$T$ behavior
of the dynamic transition line~\cite{OurFuture}: 
$\gamma_{\rm d}(\beta) \simeq \gamma_{\rm d} + x_* e^{-\beta}$
with $x_*\approx 9$ for $p=3$.
The scaling function 
behaves as $f(x)\simeq e^{-\Upsilon(\log|x|)^2/2}$ as $x\to -\infty$, and 
$f(x)\simeq f_+|x_*-x|^{-\nu(\infty)}$ as $x\to x_*>0$. 
In Fig. \ref{fig_scaling} we check the scaling hypothesis against numerical 
simulations. Remarkably, Eq.~(\ref{eq:TemperatureScaling}) 
implies a super-Arrhenius behavior at 
$\gamma_{\rm d}$: $\tau(\beta,\gamma_{\rm d}) \propto e^{\Upsilon\beta^2/2}$.

Several geometrical properties of optimal (minimum barrier) LSR's
can be determined analytically. Their {\em size} (number of sites) diverges as 
$n_{\rm bar}\sim (\gamma_{\rm d}-\gamma)^{-\nu_{\rm bar}}$, with
$\nu_{\rm bar} \approx 2.0157$ for $p=3$. They are {\em non-compact}: the
chemical distance between the root and a random site in 
an optimal LSR scales as 
$(\gamma_{\rm d}-\gamma)^{-\zeta}$, with universal exponent
$\zeta=1/2$. Further, optimal rearrangements
induce {\em dynamical correlations}, as can be understood from the following 
``experiment''. Consider a site $j\neq i$, and ask what is the minimum 
barrier $b^{(j)}_i$ to be overcome for flipping $\sigma_i$ 
{\em without} flipping $\sigma_j$. Call 
$Q^{(j)}_b = {\rm Prob}[b^{(j)}_i\ge b]$, and define the ``susceptibility''
$\chi_b = \sum_j[ Q^{(j)}_b-Q_b]$. As $T\to 0$,
$\chi_b$ describes the change in $C_i(t)$ when a pinning 
field is applied on spin $\sigma_j$ (summed over $j$),
and is a geometric analogous of the 4-point susceptibility introduced in
\cite{Chi4}. We show in Fig.~\ref{fig_pih} its critical behavior.
The main feature is a peak located
at $b_{\rm peak}\simeq -\Upsilon\log(\gamma_{\rm d}-\gamma)$ whose height
$\chi_{\rm peak}\sim (\gamma_{\rm d}-\gamma)^{-\eta}$ (with universal exponent 
$\eta=1$) estimates the number of spins $\sigma_j$ which are influential to
the dynamics of $\sigma_i$, belonging to all minimal barrier rearrangements 
for this variable.

How does our picture generalize to the $\gamma>\gamma_{\rm d}$,
$T\downarrow T_{\rm d}(\gamma)$ regime? Summarizing, we presented 
two types of results: $(i)$ dynamics proceeds by LSR's 
of size
$n\sim (\gamma_{\rm d}-\gamma)^{-\nu}$; $(ii)$ the energy barriers to be 
overcome scales as $b\sim z_{\rm act}\log n$, with 
$z_{\rm act} = \Upsilon/\nu$.
The equilibration time was estimated as
$\tau\sim e^{\beta b}\sim n^{\beta z_{\rm act}}$. At finite temperature,
the Arrhenius  argument does not  make sense any more, and one cannot 
understand slowing down in terms of activated processes.
However, we still expect that LSR~\footnote{Of course we assume that 
finite-$T$ rearrangements can indeed be properly defined. See~\cite{OurFuture} 
for a discussion.} sizes diverge as 
$n\sim (T-T_{\rm d}(\gamma))^{-\nu(\gamma)}$, and that a dynamical scaling 
relation $\tau\sim n^{z}$ holds with an universal exponent $z$.
A partial confirmation is provided by the probabilistic
argument discussed in the previous pages implying
$\tau\gtrsim (T-T_{\rm d}(\gamma))^{-1/2}$.

How is this related to the issue (B) raised in the introduction?
The depth and cooperativity of LSR diverge with  
two universal exponents $\zeta=1/2$
and $\eta=1$. This agrees with MCT calculations~\cite{BB} 
implying that such universal features of MCT are not modified
by activated processes, \emph{even} in the regime 
$e^{-\beta}\ll (\gamma_{\rm d}-\gamma)$. Other features (e.g. the 
relation $\Gamma(1-a)^{2}/\Gamma(1-2a)=\Gamma(1+b)^{2}/\Gamma(1+2b)$
between $\alpha$- and $\beta$-relaxation exponents) are indeed modified
in a  crossover region that can be experimentally relevant.
However, the asymptotic 
$T\downarrow T_{\rm d}(\gamma)$ behavior is governed by usual MCT at any 
$\gamma>\gamma_{\rm d}$. The crossover between the two regimes is ruled
by the ratio $(\gamma-\gamma_{\rm d})^{-1}/e^{\beta}$. In a more general 
context one should consider the ratio
$\xi_{\rm dyn}/\xi_{\rm therm}$, where $\xi_{\rm dyn}$ is a dynamical 
length scale as measured through 4-points correlations~\cite{BB}, and 
$\xi_{\rm therm}$ is a thermal length (distance between energy defects).

The above ideas can be applied to particle systems. In the particular
case of kinetically constrained models on Bethe lattices~\cite{RigidityAndCo}, 
we could show that the same scenario described above 
holds~\cite{OurFuture}.
A challenging direction would be to analyze  ensembles of NP-hard
decision problems (random $K$-SAT, or the $q$-coloring of random 
graphs) with a similar phase diagram~\cite{SATColoring}. 
 Finally, we 
obtained a purely geometrical description of diverging spatial structures 
at the DPT. This provides a
particularly concrete setting for discussing finite-dimensionality
effects.

We thank Leticia Cugliandolo for her interest in this work. G.S. has been
partially supported by the EU under the EVERGROW project.

\end{document}